\documentclass[a4paper,aps,superscriptaddress,nofootinbib,floatfix,epfs]{article}
\usepackage[margin=2.75cm]{geometry}
\usepackage{amsfonts,amscd,amsmath,amssymb,graphicx,graphbox,color,float,relsize,authblk}

\def\ep{\text{e}}
\def\g{\mathsf{g}}
\def\oh{\frac{1}{2}}
\def\s{\mathsf{s}}
\def\m{\mathsf{m}}

\def\n{\mathsf{n}}

\def\rv{r_\text{\tiny v}}
\def\rq{r_\text{\tiny q}}
\def\rw{r_\text{\tiny w}}
\def\fv{f_\text{\tiny v}}
\def\est{\sigma_\text{\tiny $\bot$}}
\def\estv{\sigma_\text{\tiny $\parallel$}}
\def\eeff{\sigma_\text{\tiny eff}}
\def\rmin{r_\text{\tiny min}}
\def\rmax{r_\text{\tiny max}}

\def\Qqb{\text{\tiny Q}\bar{\text{\tiny q}}}

\def\QQb{\text{\tiny Q}\bar{\text{\tiny Q}}}
\def\Q{\text{\tiny Q}}

\def\rh{r_h}

\def\r0{r_{\text{\tiny 0}}}

\begin{document}
\title{String breaking in a cold wind as seen by string models}
\author{Oleg Andreev}
 \affil{L.D. Landau Institute for Theoretical Physics, Kosygina 2, 119334 Moscow, Russia}
\affil{Arnold Sommerfeld Center for Theoretical Physics, LMU-M\"unchen, Theresienstrasse 37, 80333 M\"unchen, Germany}
\date{}
\maketitle
\begin{abstract} 
Using the gauge/string duality, we model a heavy quark-antiquark pair in a color singlet state moving through a cold medium and explore the consequences of temperature and velocity on string breaking. In doing so, we restrict to the case of two dynamical flavors. We show that the string breaking distance slowly varies with temperature and velocity away from the critical line but could fall near it. 
\end{abstract}
\vspace{-9.25 cm}
\begin{flushright}
{\small LMU-ASC 04/21 }\\
\end{flushright}
\vspace{9.25cm}


\section{Introduction}

A special sector of hadron spectroscopy is that where all valence quarks are heavy, and the light quarks being sea quarks contribute only by virtual pairs in the field surrounding the heavy quarks. It is well-known that potential models successfully elucidate the quarkonia spectrum.\footnote{For more details and references, see the review articles \cite{rev}.} In these models one assumes that at leading order in the inverse heavy quark mass the interaction between heavy quarks and antiquarks inside hadrons can be described by means of a potential. The most significant effect of sea quarks on the potential is its flattening at large quark separations. This phenomenon is called string breaking. Clearly, it should be taken seriously in any attempt to accurately model hadrons in the framework of potential models.

Apart from hadron spectroscopy, in the last decades heavy-ion collision experiments have brought new interest into the study of hadronic matter under intense conditions of temperature and density. For certain physical phenomena it is important to understand the properties of heavy mesons moving through the medium.\footnote{For a recent review, see \cite{roth} and references therein.} In this paper we consider the effects of temperature and velocity on string breaking for a heavy quark-antiquark pair in a cold medium. Unlike a hot medium in which the consequences of those on Debye screening were widely discussed in the literature \cite{ubook}, this issue has not been addressed yet.   

Although lattice gauge theory remains a basic tool for studying nonperturbative phenomena in QCD \cite{bali}, at present there are no lattice results on string breaking at finite temperature and velocity. On the other hand, the gauge/string duality \cite{ubook} provides new theoretical tools for studying strongly coupled gauge theories, and therefore may be used as an alternative way to tackle this problem, or at least to gain some insight into it. In that framework, the estimate of the string breaking distance was made in \cite{KKW} for zero temperature and later in \cite{a-strb1} for finite temperature and non-zero chemical potential. The present paper continues these studies for the case of two dynamical flavors. The rest is organized as follows. We start in section 2 by describing a basic framework for understanding and engineering string configurations. Then, using this framework, we construct the configurations describing a quark-antiquark system moving through a thermal medium and derive explicit formulas for the string breaking distance. As an application, we consider in section 3 a simple model to show how this works in practice. We conclude in section 4 with a few comments on other string configurations and the behavior near the critical line.

\section{Calculating the String Breaking Distance }
\renewcommand{\theequation}{2.\arabic{equation}}
\setcounter{equation}{0}

We begin with a rather general discussion on string breaking in the framework of $\text{AdS/QCD}$. In the next section, we will consider one simple model that illustrates these ideas. 

\subsection{Preliminaries}

First, let us set the basic framework. We deal with effective string models in five dimensions. The metric (with Euclidean signature) is taken to be of the form 

\begin{equation}\label{metric}
ds^2=\ep^{\s r^2}\frac{R^2}{r^2}\Bigl(f(r)dt^2+dx_i^2+f^{-1}(r)dr^2\Bigr)
\,.
\end{equation}
Here $t$ is  periodic with period $1/T$ and $i$ goes from $1$ to $3$. It has a boundary at $r=0$ which is $\mathbf{S}^1\times\mathbf{R}^3$. We assume that the blackening factor $f$ is a decreasing function of $r$ on the interval $[0,\rh]$ such that $f(0)=1$ on the boundary and $f(\rh)=0$ on the horizon. One can think of this geometry as a one-parameter deformation, parameterized by $\s$, of the Schwarzschild black hole on $\text{AdS}_5$ space of radius $R$. The Hawking temperature associated to the black hole is given by $T=\frac{1}{4\pi}\vert \frac{df}{dr}\vert_{r=\rh}$. In the dual description it is interpreted as the temperature of gauge theory. An important feature of this geometry in the confined phase is the existence of a soft wall which stops strings from getting deeper in the bulk.

Before we go on we need to clarify one important point. We don't require that the metric be a solution of some equations. We will study classical Nambu-Goto strings in this geometry. From the Weyl anomaly point of view, see for instance \cite{joe1}, $\beta^G_{\mu\nu}=0$ is required, but for our purposes is not all that relevant since we work at the classical level whereas the anomaly appears at the one loop level.

We consider a quark-antiquark pair moving though a thermal medium. We assume that the quarks are heavy enough such that the pair is moving uniformly, as usual in the string models \cite{argy}. But in the reality there is, of course, energy lost that makes the system time-dependent and, as a consequence, the analysis becomes extremely tedious and complicated. It is convenient to choose the rest frame so that the medium takes on the role of a wind blowing on the static pair. To describe this, one first Wick rotates the background metric to Minkowski form and then boosts to the rest frame of the pair \cite{ubook}. For the wind of velocity $v$ in the $z$-direction, a simple calculation shows that\footnote{We suppress the primes as we will always be interested in the pair rest (primed) frame.}   

\begin{equation}\label{metricv}
ds^2=
\ep^{\s r^2}\frac{R^2}{r^2}
\Bigl(-\fv dt^2+2v \gamma^2(1-f)dt dz+dx^2+dy^2+\gamma^2(1-fv^2)dz^2
+
 f^{-1}dr^2\Bigr)
\,,
\end{equation}
with $\fv(r)=\gamma^2(f-v^2)$ and $\gamma=1/\sqrt{1-v^2}$. This metric has an induced horizon at $r = \rv$, where the function $\fv(r)$ vanishes. Since $f$ is a decreasing function of $r$, it holds that $\rv<\rh$ for all $v\not =0$.

The string in question is a Nambu-Goto string governed by the action 

\begin{equation}\label{NG}
S=-\frac{1}{2\pi\alpha'}\int d^2\xi\,\sqrt{-\det\gamma_{\alpha\beta}}
\,.
\end{equation}
Here $\alpha'$ is a string constant, $(\xi_1,\xi_2)$ are worldsheet coordinates, and $\gamma_{\alpha\beta}$ is an induced metric on the worldsheet. 

To take account of light quarks at string endpoints, we introduce a background scalar field. This is not a new idea for dealing with light flavors in AdS/QCD, but the earlier attempts were primarily based on the space-time formalism, i.e. on the use of effective five-dimensional field theories on $\text{AdS}$ space \cite{son}. In contrast, we do so in the worldsheet formalism. Since we wish to mimic a $SU(3)$ gauge theory with two light dynamical flavors in the case when the $u$ and $d$ quarks have equal masses, one boundary term $S_{\text{q}}=-\int d\tau e\,\text{T}$ is needed.  Here $\tau$ is a coordinate on a worldsheet boundary, $e$ is a boundary metric, and $\text{T}$ is a background scalar which could usually be interpreted as an open string tachyon background.\footnote{In the current context, such a tachyon signals an instability of a fundamental string rather than a D-brane.} For the reason of simplicity, we take a constant background field $\text{T}=\text{T}_0$. The important fact for us will be that the action written in the static gauge $t=\xi^1$ is of the form \cite{a-strb1}  

\begin{equation}\label{Sq}
S_{\text q}=-\m\int dt \sqrt{\fv}\,\frac{\ep^{\frac{\s}{2}r^2}}{r}
\,,
\end{equation}
with $\m=R{\text T}_0$. Clearly, it is an action for a particle of mass ${\text T}_0$ at rest.

\subsection{Connected Configurations}

\begin{figure}[htbp]
\centering
\includegraphics[width=13.5cm]{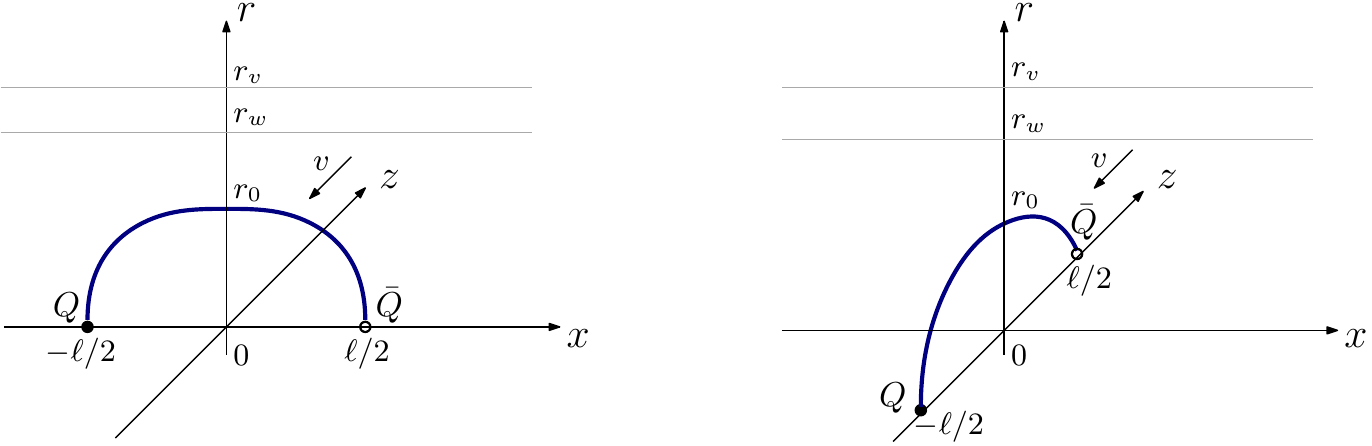}
\caption{{\small Connected string configurations. The strings are in the ground state. The horizontal lines at $r=\rv$ represent the induced horizons and those at $\rw$ the soft walls. The heavy quarks are separated by distance $\ell$. The wind velocity is perpendicular to the quark-antiquark axis (left) and parallel to it (right).}}
\label{conconfs}
\end{figure}

 We now want to consider string configurations which are static in the rest frame of the pair. In each of these configurations, the heavy quark sources $Q$ are placed on the boundary and connected by a string. We concentrate on two special cases, where the wind velocity is either perpendicular or parallel to the quark-antiquark axis. The corresponding configurations are sketched in figure \ref{conconfs}.\footnote{In the context of AdS/CFT (without the soft wall), the analogous configurations were discussed in \cite{nats}.} 
For a reason that will become clear shortly, we assume that the soft wall is closer to the boundary than the induced horizon. The total action is the sum of the Nambu-Goto action and those of two point-like quark sources. The last are formally divergent as the rest energies of infinitely heavy sources. In the real world the charm and bottom quarks are heavy, but not infinitely so. To take this into account, we include the finite rest energies in the definition of a normalization constant $c$ defined below. 

Let us consider first the transverse case, as shown in the left panel of the figure. The analysis proceeds along the lines of \cite{a-mD} and involves only one new ingredient related to the large $\ell$ behavior of the energy. So, we choose the static gauge $\xi_1 = t$ and $\xi_2 = x$, and consider $r$ as a function of $x$ only.\footnote{Of course, a smooth string can bend in the wind, but then it is non-static or not energetically favorable. We give an example of the latter in section 4.} Then the expression \eqref{NG} for the Nambu-Goto action becomes 

\begin{equation}\label{NG-per}
	S=-\g\int dt \int_{-\ell/2}^{\ell/2} dx\,
	\est\sqrt{1+f^{-1}r'{}^2}
	\,,
	\qquad\text{with}\qquad
	\est(r)=\frac{\ep^{\s r^2}}{r^2}\sqrt{\fv}
	\,.
	\end{equation}
Here $\g=\frac{R^2}{2\pi\alpha'}$ and $r'=\frac{\partial r}{\partial x}$. The boundary conditions at the string endpoints are $r(\pm\ell/2)=0$. Since the Lagrangian is explicitly independent of $x$, there exists a conserved quantity which is a first integral of the equation of motion 

\begin{equation}\label{I-per}
I=\frac{\est}{\sqrt{1+f^{-1}r'{}^2}}
\,.
\end{equation}
As long as $r<\rw$, $r(x)$ is a smooth function. On symmetry grounds, $r'(0)=0$ that allows us to set $I=\est(\r0)$ with $\r0=r(0)$. The string length along the $x$-axis can be found by integrating \eqref{I-per}, with the result

\begin{equation}\label{l-per}
\ell=2\int_0^{\r0} \frac{dr}{\sqrt{f}}\, 
\biggl[\,
\frac{\est^2}{I^2}-1\,
\biggr]^{-\oh}
\,.
\end{equation}
The factor of $2$ comes from the reflectional symmetry of the configuration.

Given a string configuration (solution), one can compute its energy. Since the string is static, the energy is simply related to the Lagrangian. The only subtlety is that the integral is divergent at $r=0$ due to the factor $r^{-2}$ in $\est$. We regularize it by imposing a cutoff $\epsilon$. Thus, the regularized expression takes the form

\begin{equation}\label{E-perR}
E_{\QQb}^R=2\g\int_{\epsilon}^{\r0} dr\,\frac{\est}{\sqrt{f}}
\biggl[\,1-\frac{I^2}{\est^2}\,\biggr]^{-\oh}
\,.
\end{equation}
Its $\epsilon$-expansion is 

\begin{equation}\label{E-perR2}
E_{\QQb}^R=\frac{2\g}{\epsilon}+E_{\QQb}+O(\epsilon)
\,.
\end{equation}
Subtracting the $\frac{1}{\epsilon}$-term and letting $\epsilon = 0$, we get a finite result

\begin{equation}\label{E-per}
E_{\QQb}=2\g\int_{0}^{\r0} dr\biggl(\frac{\est}{\sqrt{f}}
\biggl[\,1-\frac{I^2}{\est^2}\,\biggr]^{-\oh}-\frac{1}{r^2}\biggr)\,\,-\frac{2\g}{\r0}+2c
\,,
\end{equation}
with a renormalization constant $c$. 

\begin{figure*}[htbp]
\centering
\includegraphics[width=6.5cm]{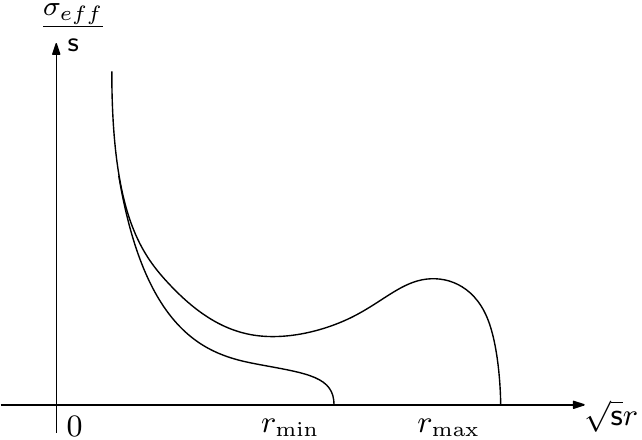}
\caption{{\small Schematic representation of the effective string tension in two different regimes: confinement (upper curve) and deconfinement (lower curve). The upper curve has a local minimum at $r=\rmin$ and a local maximum at $r=\rmax$. The effective string tension vanishes at $r=\rv$.}}
\label{Est}
\end{figure*}

In studying the long distance behavior of $E_{\QQb}$, it is helpful to consider an effective string tension $\eeff$ \cite{az3}. The explicit formula for it follows from the action \eqref{NG-per} evaluated at a constant solution $r=const$. So, we have $\eeff=\est$. There is a simple but important fact: the form of $\est$, as a function of $r$, is temperature and velocity dependent. Indeed, for small $T$ and $v$, where $\fv\approx 1$, $\est$ has a local minimum near $r=1/\sqrt{\s}$ defined by the warp factor. On the other hand, for large enough $T$ and $v$, where $\rv\ll 1/\sqrt{\s}$, there is no local minimum as both factors are decreasing functions in the interval $[0,\rv]$. We illustrate this in figure \ref{Est}.
If the local minimum exists, then a string cannot get deeper than $\rmin$ in the $r$-direction because a repulsive force prevents it from doing so. Thus, the soft wall is located at $r = \rmin$ so that $\rw\equiv\rmin$. From this, it follows that $\rw<\rv$. The large distance behavior of the string is completely determined by the wall. The leading term is obtained by setting $r=\rmin$ in \eqref{NG-per}. Explicitly, 

\begin{equation}\label{Elarge-per}
	E_{\QQb}=\sigma\ell+O(1)
	\,,\quad\text{with}\quad 
	\sigma=\g\est(\rmin)
\,.
\end{equation}
In the absence of the local minimum, the string can reach the induced horizon, where the effective tension vanishes. As a result, the linear term drops out in the expansion \eqref{Elarge-per}. The first type of behavior can be interpreted as the phase of confinement, with the physical string tension $\sigma$, while the second as the phase of deconfinement. In the $Tv$-plane, the boundary between the phases is determined from the equation $\rmin=\rmax$. 

Thus, in the confined phase the energy of the connected configuration is given by the parametric equations \eqref{l-per} and \eqref{E-per}, with $\r0\in[0,\rmin]$. Let us also note that at $v=0$ the formulas are reduced to those of \cite{az3}.

To find the constant term in the asymptotic expansion of $E_{\QQb}$, consider

\begin{equation}\label{E-large00}
	E_{\QQb}-\sigma\ell=2\g\int_{0}^{\r0} dr\biggl(\frac{\est}{\sqrt{f}}
\biggl[\,1-\frac{I^2}{\est^2}\,\biggr]^{-\oh}-\frac{\est(\rmin)}{\sqrt{f}}\biggl[\,\frac{\est^2}{I^2}-1\,\biggr]^{-\oh}
-\frac{1}{r^2}\biggr)\,\,
-\frac{2\g}{\r0}+2c
	\,.
\end{equation}
After taking the limit $\r0\rightarrow\rmin$, we find  

\begin{equation}\label{E-large1}
	E_{\QQb}=\sigma\ell+2\g\int_{0}^{\rmin} dr\biggl(\frac{\est}{\sqrt{f}}
\biggl[\,1-\frac{\est^2(\rmin)}{\est^2}\,\biggr]^{\oh}-\frac{1}{r^2}\biggr)\,\,
-\frac{2\g}{\rmin}+2c+o(1)
	\,.
\end{equation}
This is the formula we will use to estimate the string breaking distance. 

Now consider the longitudinal case, as shown in the figure \ref{conconfs}. In that case it is convenient to choose the static gauge $\xi_1=t$ and $\xi_2=z$, and consider $r$ as a function of $z$. The Nambu-Goto action is then written as 

\begin{equation}\label{NGp}
	S=-\g\int dt \int_{-\ell/2}^{\ell/2} dz\,
	\estv\sqrt{1+\frac{\fv}{f^2}r'{}^2}
	\,\,,
	\qquad\text{with}\qquad
\estv(r)=\frac{\ep^{\s r^2}}{r^2}\sqrt{f}
\,.
	\end{equation}
Here $r'=\frac{\partial r}{\partial z}$. The boundary conditions at the string endpoints are given again by $r(\pm\ell/2)=0$. A first integral of the equation of motion is 

\begin{equation}\label{I-par}
I=\frac{\estv}{\sqrt{1+\frac{\fv}{f^2}r'{}^2}}
\,.
\end{equation}
The reasoning is analogous to that in the previous case.

If $r\leq\rw$, then $r(z)$ is smooth. The subsequent analysis therefore can proceed in close parallel to the transverse case. Again, it is convenient to set $I=\estv(\r0)$, where $\r0$ is the turning point shown in figure \ref{conconfs}. It follows from \eqref{I-par} that the integral over $[-\ell/2,\ell/2]$ of $dz$ is equal to 

\begin{equation}\label{l-par}
\ell=2\int_0^{\r0} \frac{dr}{f}\,\sqrt{\fv} 
\biggl[\,
\frac{\estv^2}{I^2}-1\,
\biggr]^{-\oh}
\,.
\end{equation}
To compute the energy, we first reduce the integral over $z$ in eqn.\eqref{NGp} to that over $r$. After that, we regularize it by imposing a short distance cutoff $\epsilon$. So,

\begin{equation}\label{E-par00}
E_{\QQb}^R=2\g\int_{\epsilon}^{\r0} dr
\frac{\est}{\sqrt{f}}
\biggl[\,1-\frac{I^2}{\estv^2}\,\biggr]^{-\oh}
\,.
\end{equation}
Near $\epsilon = 0$, the expansion of $E_{\QQb}^R$ is of the form \eqref{E-perR2}. Subtracting the $\frac{1}{\epsilon}$-term and letting $\epsilon = 0$ yields

\begin{equation}\label{E-par}
E_{\QQb}=2\g\int_{0}^{\r0} dr\biggl(\frac{\est}{\sqrt{f}}
\biggl[\,1-\frac{I^2}{\estv^2}\,\biggr]^{-\oh}-\frac{1}{r^2}\biggr)\,\,-\frac{2\g}{\r0}+2c
\,.
\end{equation}
Here $c$ is the same renormalization constant as in \eqref{E-per}. Thus, the energy of the configuration is written parametrically as $E_{\QQb}=E_{\QQb}(\r0)$ and $\ell=\ell(\r0)$, with the parameter varying from $0$ to $\rmin$.

Once the parametric equations are given, it is not difficult to analyze the long distance behavior of $E_{\QQb}$. From \eqref{NGp}, it follows that the effective string tension is $\eeff=\estv$ which is a special case of $\est$ with $v=0$. The long distance behavior is governed by the soft wall. Substituting $r=\rmin$ in $S$, we get 

\begin{equation}\label{Elarge-par}
	E_{\QQb}=\sigma\ell+O(1)
	\,,\quad\text{with}\quad 
	\sigma=\g\estv(\rmin)
\,.
\end{equation}
It is noteworthy that the longitudinal wind does not affect the string tension. The next term in the expansion is determined by writing 

\begin{equation}\label{Elarge-par2}
	E_{\QQb}-\sigma\ell=2\g\int_{0}^{\r0} dr\biggl(\frac{\est}{\sqrt{f}}
\biggl[\,1-\frac{I^2}{\estv^2}\,\biggr]^{-\oh}
-
\frac{\estv(\rmin)\sqrt{\fv}}{f}\biggl[\,\frac{\estv^2}{I^2}-1\,\biggr]^{-\oh}
-\frac{1}{r^2}\biggr)\,\,
-\frac{2\g}{\r0}+2c
	\,,
\end{equation}
and then taking the limit $\r0\rightarrow\rmin$. So, we have

\begin{equation}\label{Elarge-par3}
	E_{\QQb}=\sigma\ell+2\g\int_{0}^{\rmin} dr\biggl(\frac{\est}{\sqrt{f}}
\biggl[\,1-\frac{\estv^2(\rmin)}{\estv^2}\,\biggr]^{\oh}-\frac{1}{r^2}\biggr)\,\,
-\frac{2\g}{\rmin}+2c+o(1)
	\,,
\end{equation}
which is the analog of Eq.\eqref{E-large1}. 

\subsection{Disconnected Configurations and String Breaking}

\begin{figure}[htbp]
\centering
\includegraphics[width=13cm]{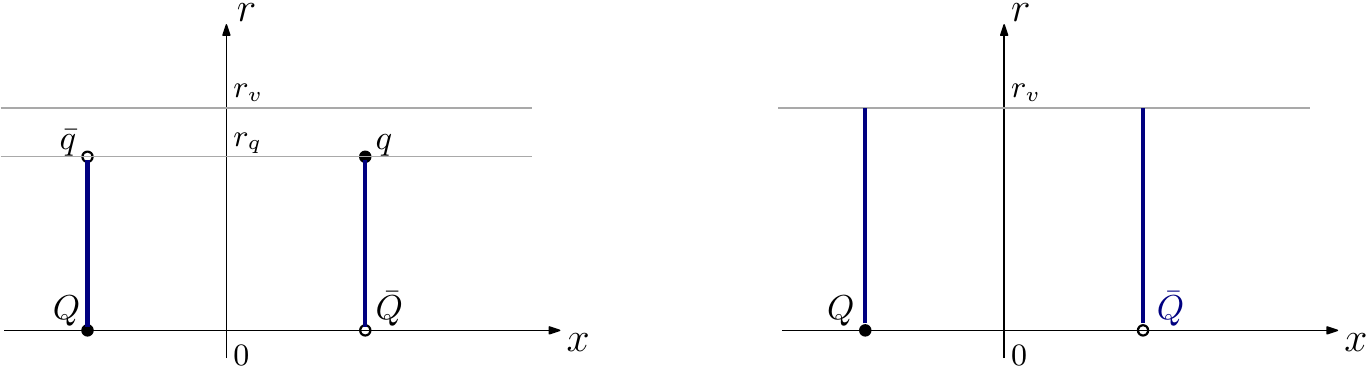}
\caption{{\small Disconnected string configurations. The light quark (antiquark) is denoted by $q\,(\bar q)$ and located at $r=\rq$.}}
\label{disc}
\end{figure}

To get further, consider the disconnected string configurations shown in figure \ref{disc} which dominate at large quark separation. The first configuration is interpreted 
as a pair of non-interacting heavy-light mesons. It dominates at low temperatures, where a string breaks via the production of a $q\bar q$ virtual pair in a strong chromoelectric field, namely $Q\bar Q\rightarrow Q\bar q+q\bar Q$. As temperature approaches the critical value, the thermal fluctuations become strong enough to destroy a string. In that case one has $Q\bar Q\rightarrow Q+\bar Q$, with the free quarks described by the second configuration.

We start our analysis with the first configuration. In this case, the action has in addition to the standard Nambu-Goto actions of the fundamental strings, contributions arising from the light quarks. It is thus $S=\sum_{i=1}^2 S_{\text{\tiny NG}}+S_{\text{q}}$. We choose the static gauge $\xi_1=t$  and $\xi_2=r$, and consider $x$ as a function of $r$. The action is  then 

\begin{equation}\label{NG-Q}
S=-2\g\int dt \int_0^{\rq} dr\,\est
\sqrt{x'{}^2+f^{-1}}\,\,
- 
2\g\n\int dt \sqrt{\fv}\,\frac{\ep^{\frac{\s}{2}\rq^2}}{\rq}
	\,,
\end{equation}
with $\n=\frac{\m}{\g}$ and $x'=\frac{\partial x}{\partial r}$. It is obvious that $x(r)=const$ is a solution to the equation of motion that represents a straight string stretched between the boundary and light quark in the bulk, as shown in the figure above. Clearly, this solution corresponds to the lowest string energy. 

Varying $S$ with respect to $r_q$ yields 

\begin{equation}\label{fbq}
\ep^{\frac{q}{2}}+\n\sqrt{f}\Bigl(q-1+q\frac{\partial\ln\fv}{\partial q}\Bigl)=0
\,.
\end{equation}
Here $q=\s\rq^2$. This is nothing else but a force balance equation that determines the light quark position on the $r$-axis. At $v=0$, it reduces to the equation derived in \cite{a-strb1}. 

The energy of the configuration is a sum of two equal terms, each of which is interpreted as an energy of a heavy-light meson. The latter is obtained by evaluating the Lagrangian on the solution $x=const$. We have

\begin{equation}\label{EQqb00}
E_{\Qqb}=\g\int_0^{\rq}\frac{dr}{\sqrt{f}}\est
+
\g\n\sqrt{\fv}\,\frac{\ep^{\frac{\s}{2}\rq^2}}{\rq}
\,.
\end{equation}
The integral is divergent at $r=0$. We regularize it by imposing the short-distance cutoff $\epsilon$. The regularized expression behaves for $\epsilon\rightarrow 0$ as

\begin{equation}\label{EQqb1}
E_{\Qqb}^{\text{\tiny R}}
=\frac{\g}{\epsilon}+E_{\Qqb}+O(\epsilon)
\,.
\end{equation}
After subtracting out the $\frac{1}{\epsilon}$-term, we get 

\begin{equation}\label{EQqb}
E_{\Qqb}=\g\int_0^{\rq}dr\biggl(\frac{\est}{\sqrt{f}}-\frac{1}{r^2}
\biggr)\,\,
-
\frac{\g}{\rq}
+
\g\n\sqrt{\fv}\,\frac{\ep^{\frac{\s}{2}\rq^2}}{\rq}
+c
\,,
\end{equation}
with the same normalization constant as before. Clearly, the total energy is twice the value in \eqref{EQqb}. Note that at $v=0$ the formulas reduce to those of \cite{a-strb1}.

The second configuration can be analyzed similarly. But from the formula \eqref{EQqb}, it is straightforward to obtain the desired result by letting $\n=0$ and replacing $\rq$ by $\rv$. So, 
 
 \begin{equation}\label{EQ}
E_{\Q}=\g\int_0^{\rv}dr\biggl(\frac{\est}{\sqrt{f}}-\frac{1}{r^2}
\biggr)\,\,
-
\frac{\g}{\rv}
+c
\,.
\end{equation}
Again, the total energy is twice this value. In the limit $v\rightarrow 0$ the integral can be evaluated with standard techniques. The result is written in terms of the imaginary error function \cite{a-pol}.

Since we are primarily interested in string breaking due to the light quarks, we have to consider a region of the parameter space where the first configuration is energetically favorable. This is equivalent to the requirement that the allowed values of $v$ and $T$ are restricted by the condition $E_{\Qqb}\leq E_{\Q}$. With the help of the formulas \eqref{EQqb1} and \eqref{EQ}, the inequality can be written as 

\begin{equation}\label{inequality}
	\n\sqrt{\fv}\,\frac{\ep^{\frac{\s}{2}\rq^2}}{\rq}
\leq
\int_{\rq}^{\rv}dr \frac{\est}{\sqrt{f}}\,
\,.
\end{equation}
For positive $\n$ values, it makes sense only if $\rq <\rv$. 

We are now ready to compute a characteristic scale of string breaking. Like in \cite{drum}, we take 

\begin{equation}\label{deflc}
E_{\QQb}(\boldsymbol{\ell}_{\QQb})=2E_{\Qqb}
\end{equation}
as a definition and call $\boldsymbol{\ell}_{\QQb}$ the string breaking distance.\footnote{Here we added a subscript $"\QQb\,"$ to indicate that it refers to the $Q\bar Q$ system. In general, $\boldsymbol{\ell}_{\QQb}$ can be different from that in the $QQQ$ system \cite{a-3Qb}.} In the dual formulation, this definition has a clear meaning such as a separation scale which separates the two string configurations from each other. The first is energetically favorable at small quark separation, while the second at large. An important point is that for large $\ell$, where a string is expected to break down, the energy of the connected configuration is well approximated by a linear function. If so, then solving the equation \eqref{deflc} is simple. Using \eqref{E-large1}, \eqref{Elarge-par3} and \eqref{EQqb}, one can show that in the transverse case

\begin{equation}\label{lc-per}
\boldsymbol{\ell}_{\QQb}^{\bot}=\frac{2}{\est(\rmin)}
\biggl(\n\sqrt{\fv}\,\frac{\ep^{\frac{\s}{2}\rq^2}}{\rq}
+
\int_0^{\rmin}dr\frac{\est}{\sqrt{f}}\Bigl(1-\Bigl[1-\frac{\est^2(\rmin)}{\est^2}\Bigr]^{\oh}\Bigl)
-
\int_{\rq}^{\rmin}dr\frac{\est}{\sqrt{f}}
\biggr)
	\,,
\end{equation}
whereas in the longitudinal case

\begin{equation}\label{lc-par}
\boldsymbol{\ell}_{\QQb}^{\parallel}=\frac{2}{\estv(\rmin)}
\biggl(\n\sqrt{\fv}\,\frac{\ep^{\frac{\s}{2}\rq^2}}{\rq}
+ 
\int_0^{\rmin}dr\frac{\est}{\sqrt{f}}
\Bigl(1-\Bigl[1-\frac{\estv^2(\rmin)}{\estv^2}\Bigr]^{\oh}\Bigl)
-
\int_{\rq}^{\rmin}dr\frac{\est}{\sqrt{f}}
\biggr)
	\,.
\end{equation}
The parameters $\rmin$ which enter into the right hand side of the equations are different, but we omit the subscripts when it is clear from the context. In the limit $T\rightarrow 0$, in which $f\rightarrow 1$, both these expressions give the expected result \cite{a-strb1}

\begin{equation}\label{Lc0}
	\boldsymbol{\ell}_{\QQb}^{(0)}=\frac{2}{\ep\sqrt{\s}}
\Bigl(
{\cal Q}(q)+\n\frac{\ep^{\oh q}}{\sqrt{q}}+I_0
\Bigr)
\,,
\end{equation}
with ${\cal Q}(x)=\sqrt{\pi}\text{erfi}(\sqrt{x})-\frac{\ep^x}{\sqrt{x}}$ and $I_0=\int_0^1\frac{du}{u^2}\Bigl(1+u^2-\ep^{u^2}\Bigl[1-u^4\ep^{2(1-u^2)}\Bigr]^{\frac{1}{2}}\Bigr)$. Numerically, $I_0\approx0.751$.

\section{A Sample Model}
\renewcommand{\theequation}{3.\arabic{equation}}
\setcounter{equation}{0}

So far our discussion was general. To illustrate the ideas, we will now specify the blackening factor to one of those explored in the literature \cite{PC}. It is \footnote{Such a factor was suggested in \cite{az2} for modeling the thermal properties of gauge theories in the soft wall (metric) model.} 

\begin{equation}\label{f-az}
f=1-\frac{r^4}{\rh^4}
\,.
\end{equation}
Thus the background geometry can be thought of as the slightly deformed Schwarzschild black hole in $\text{AdS}_5$ space, with a deformation parameter $\s$. The reasons for choosing this particular factor as an example are twofold: (1) it provides the results consistent with the lattice QCD and phenomenology, in particular for the Debye screening mass in QCD with two flavors \cite{a-D}, and (2) it makes computations simpler and also enables most of the results to be obtained analytically.

With this choice of the blackening factor, the Hawking temperature is simply

\begin{equation}\label{Taz2}
T=\frac{1}{\pi\rh}
\,,
\end{equation}
and the expressions for the induced blackening factor and horizon are

\begin{equation}\label{frv}
\fv=1-\frac{r^4}{\rv^4}
\,,
\qquad
\rv=\rh\sqrt[4]{1-v^2}
\,.
\end{equation}
Then a simple analysis shows that the function $\est (r)$ has local extrema at 

\begin{equation}\label{roots}
\rmin=\rv\sqrt{\frac{2}{\sqrt{3}}\sin\phi}
\,,\qquad
\rmax=\rv\sqrt{\frac{2}{\sqrt{3}}\sin
\Bigl(\frac{\pi}{3}-\phi\Bigr)}
\,,
\end{equation}
where $\phi=\frac{1}{3}\arcsin\gamma\frac{T^2}{T_{pc}^2}$ and $T_{pc}=\frac{1}{\pi}\sqrt{\frac{2\s}{3\sqrt{3}}}$. Each of these formulas has an immediate extension to the case of $\estv$. One just needs to set $v=0$. Note that at a given temperature $\rmin$ increases with velocity. This implies that the soft wall is closer to the boundary in the longitudinal case than in the transverse one.

\begin{figure}[htbp]
\centering
$\vcenter{\hbox{\includegraphics[width=5.5cm]{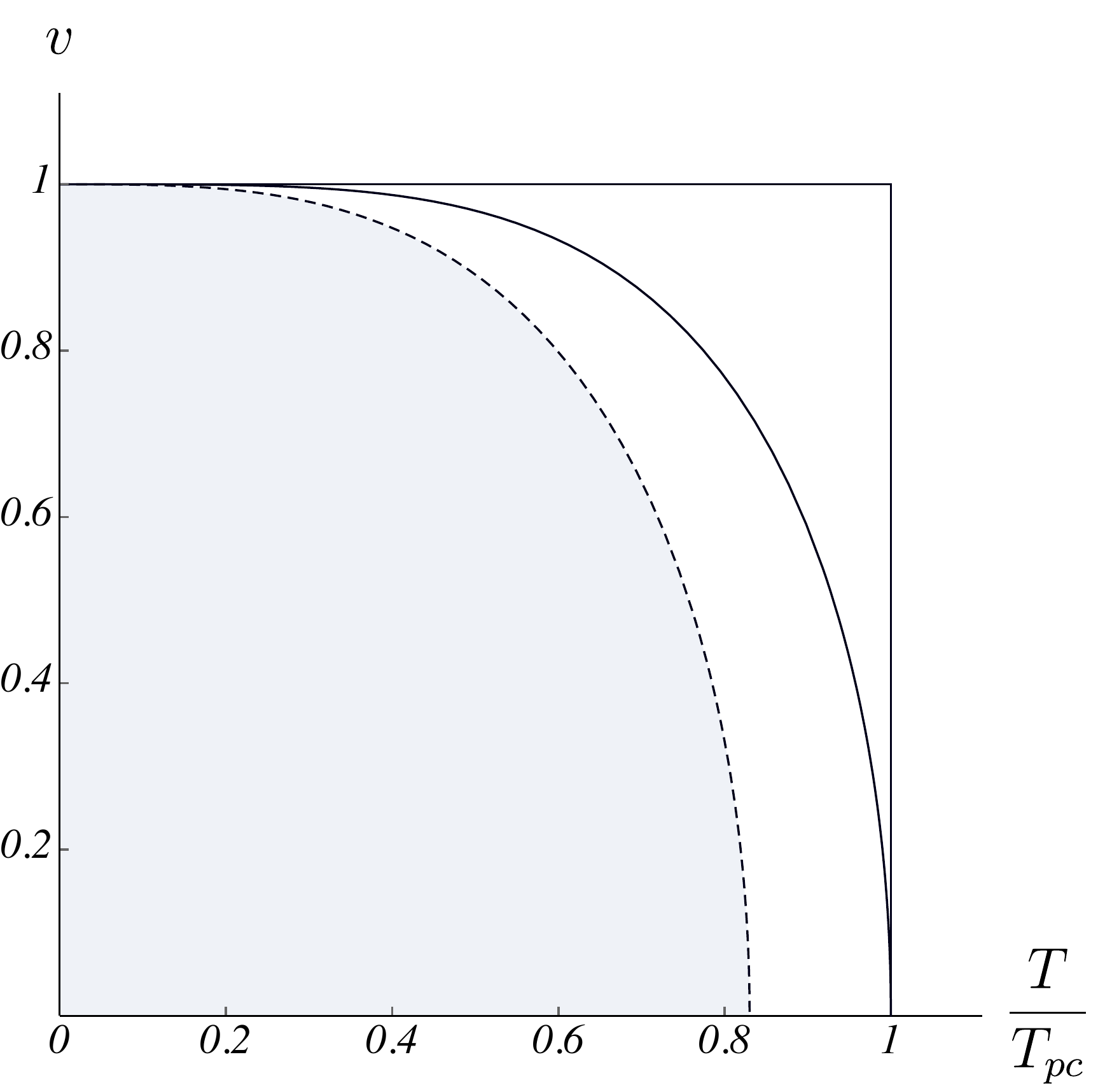}}}$
\hspace{1.5cm}
$\vcenter{\hbox{\includegraphics[width=6.2cm]{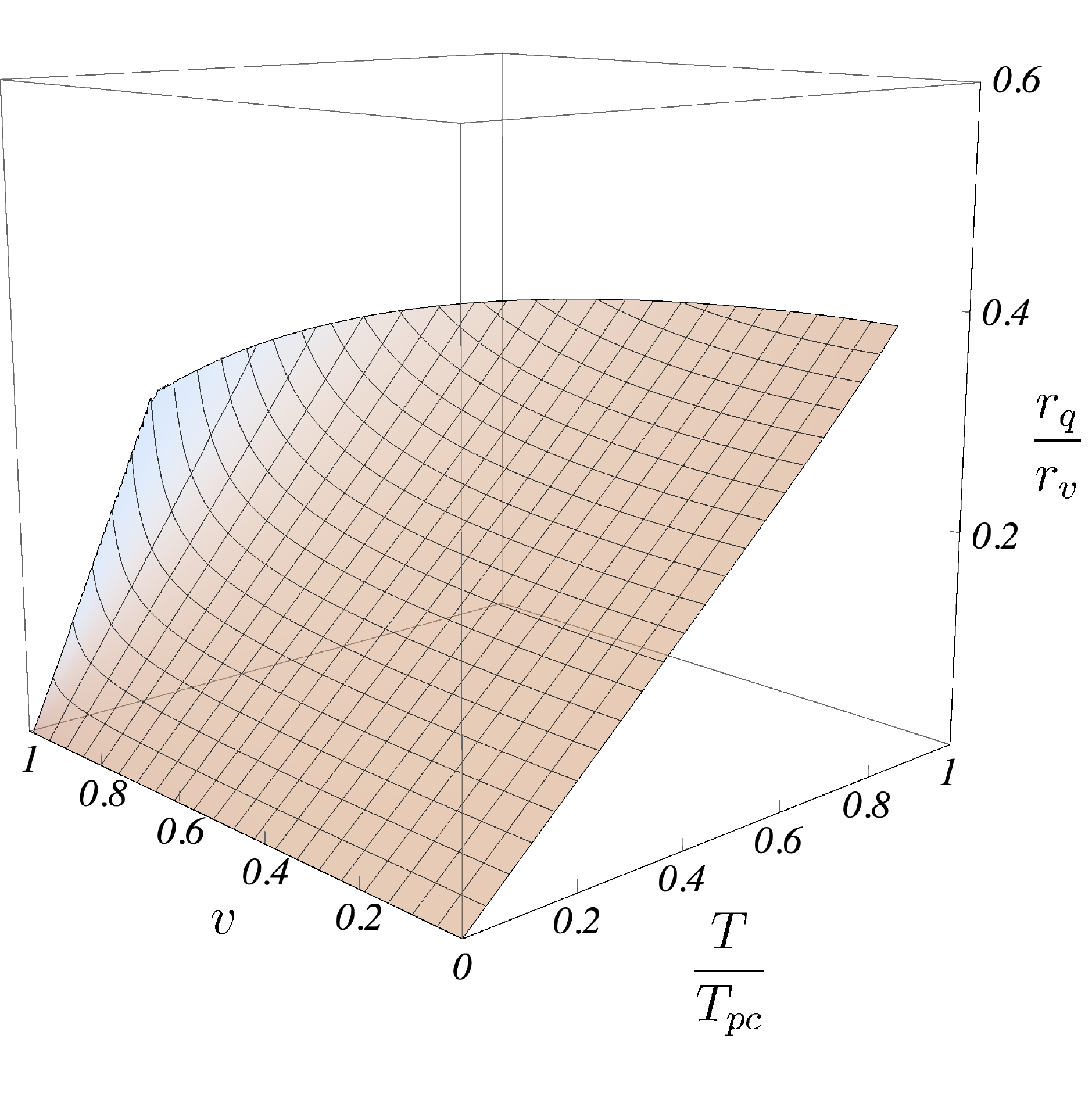}}}$
\caption{{\small Left: The shaded region is the domain of allowed values of $T$ and $v$. The solid and dashed curves correspond to the upper bounds given in equations \eqref{conf} and \eqref{supp}. Right: $\frac{\rq}{\rv}$ ratio as a function of $T$ and $v$. Here and later, for computational efficiency, we further restrict the domain of allowed values of $T$ and $v$ by setting $v\leq 0.999$.}}
\label{q}
\end{figure}

The critical line is determined by setting $\rmin=\rmax$. This gives $\phi_c=\frac{\pi}{6}$ and hence $\gamma\frac{T^2}{T_{pc}^2}=1$. Accordingly, the condition for the linear growth of $E_{\QQb}$ at large $\ell$ can be written as 

\begin{equation}\label{conf}
\frac{T}{T_{pc}}\leq (1-v^2)^{\frac{1}{4}}
\,
\end{equation}
in the transverse case and 
\begin{equation}\label{conf1}
\frac{T}{T_{pc}}\leq 1
\,
\end{equation}
in the longitudinal case. But this is not the whole story because one has to be sure  that the dominant contribution to the ground state energy at large quark separations comes from a pair of non-interacting heavy-light mesons. This condition is described by the inequality  \eqref{inequality}. Since we can't solve it analytically, we solve it numerically. In doing so, we use the parameter set suggested in \cite{a-strb1}. It is mainly a result of fitting the lattice QCD data to the string model we are considering.  The value of $\s$ is fixed from the slope of the Regge trajectory for the $\rho(n)$ mesons. As a result, we have $\s=0.450\,\text{GeV}^2$ \cite{a-q2}. Then, the value of $\g$ is fixed by fitting the value of the physical string tension at $T=0$, which is $\sigma=\ep\g\s$, to its value in \cite{bulava}. This results in $\g=0.176$. Finally, the parameter $\n$ is adjusted to reproduce the lattice result for the string breaking distance at $T=0$. With $\boldsymbol{\ell}_{\QQb}^{(0)}=1.22\,\text{fm}$ \cite{bulava}, this gives $\n=3.057$. For these parameter values, the solution to \eqref{inequality} is given to a good approximation by  

\begin{equation}\label{supp}
\frac{T}{T_{pc}}\lessapprox 0.830\,(1-v^2)^{0.322}
\,.
\end{equation}
It is what determines the domain for $T$ and $v$.\footnote{Note that $T_{pc}=132\,\text{MeV}$ at $\s=0.45\,\text{GeV}^2$.} In the process we checked the consistency condition $\rq<\rv$ by solving numerically the force balance equation \eqref{fbq} and comparing the result with the explicit expression \eqref{frv} for the induced horizon.\footnote{In a similar way one can check that $\rq<\rmin$.} For completeness, we illustrate these in figure \ref{q}. 

\begin{figure}[htbp]
\centering
\includegraphics[width=7cm]{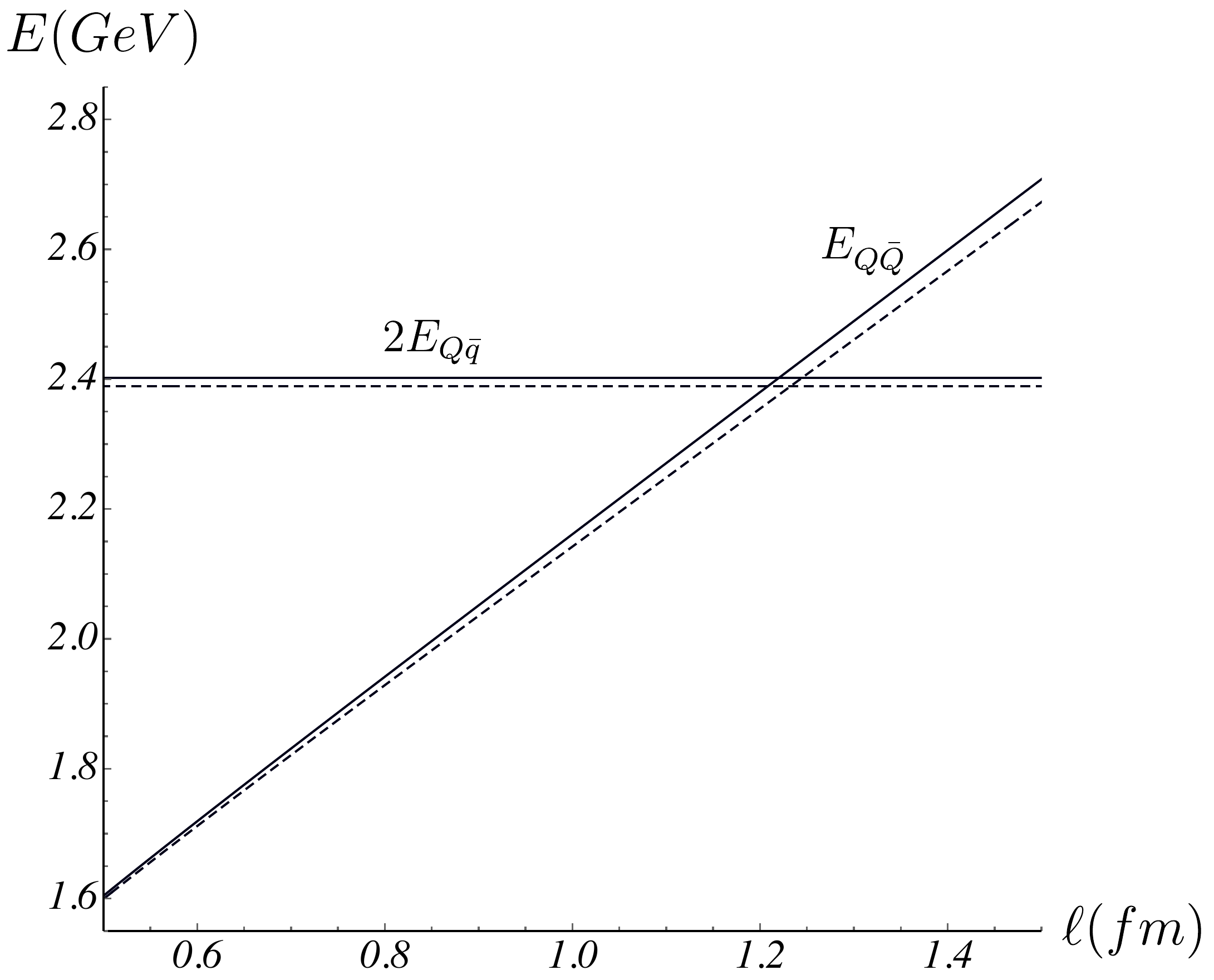}
\hspace{1.25cm}
\includegraphics[width=7cm]{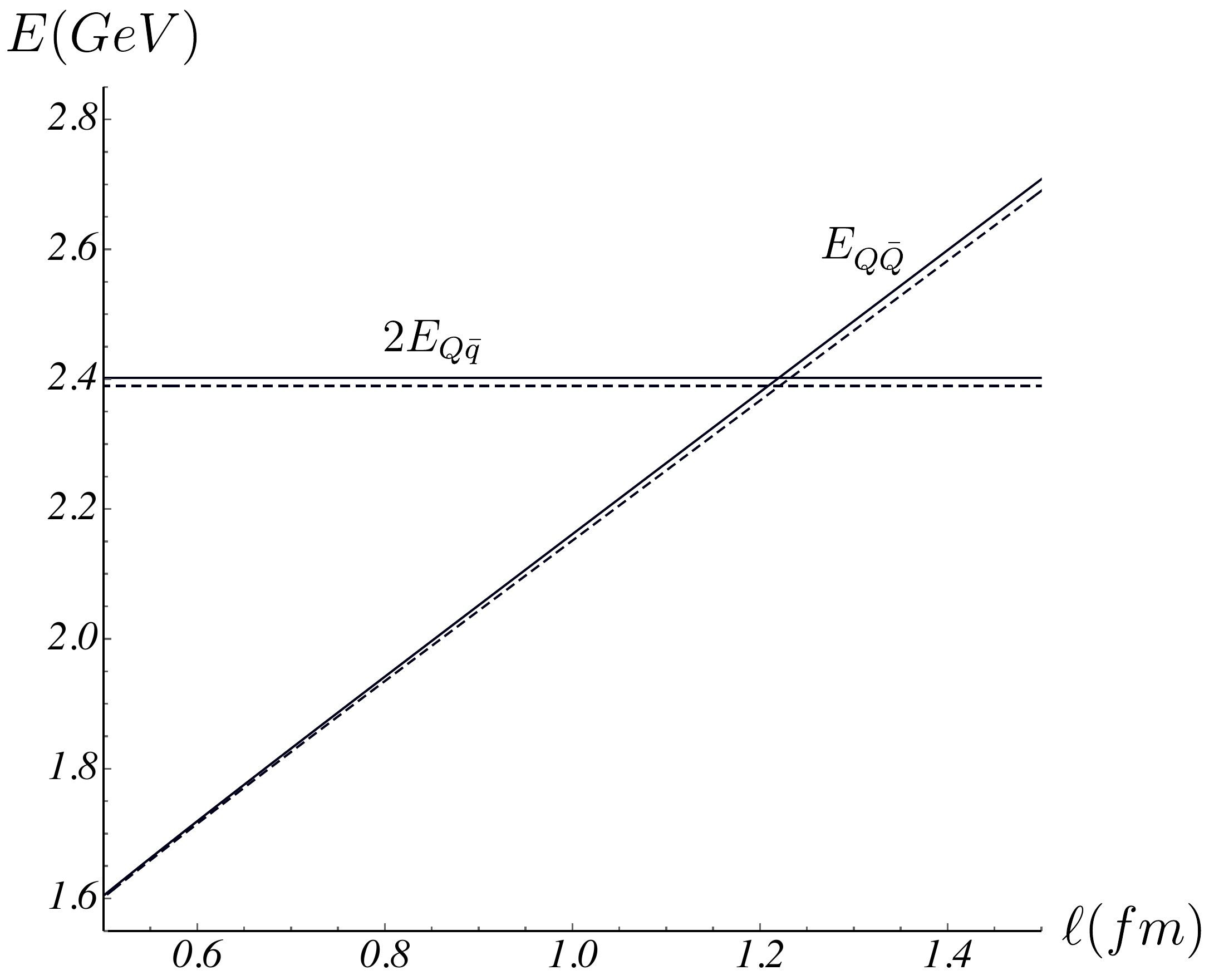}
\caption{{\small The energies $E_{\QQb}$ and $2E_{\Qqb}$ shown for the transverse and longitudinal cases, respectively. In $(\frac{T}{T_{pc}},v)$ notation, the solid lines correspond to $(0.01,0.01)$ and the dashed to $(0.66,0.70)$. We set $c=0.623\,\text{GeV}$.}}
\label{Elarge}
\end{figure}

\begin{figure}[htbp]
\centering
$\vcenter{\hbox{\includegraphics[width=7.25cm]{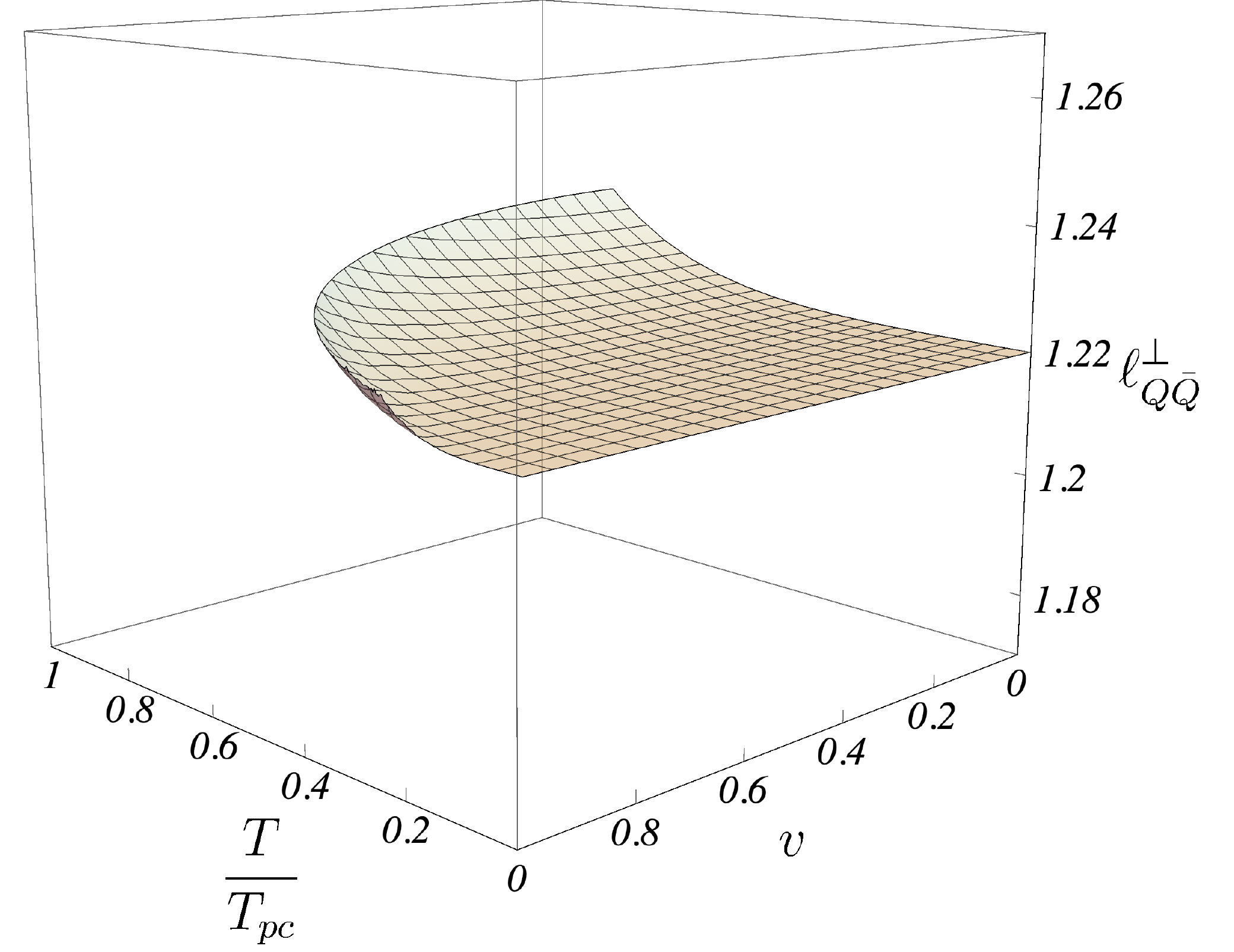}}}$
\hspace{0.65cm}
$\vcenter{\hbox{\includegraphics[width=7.05cm]{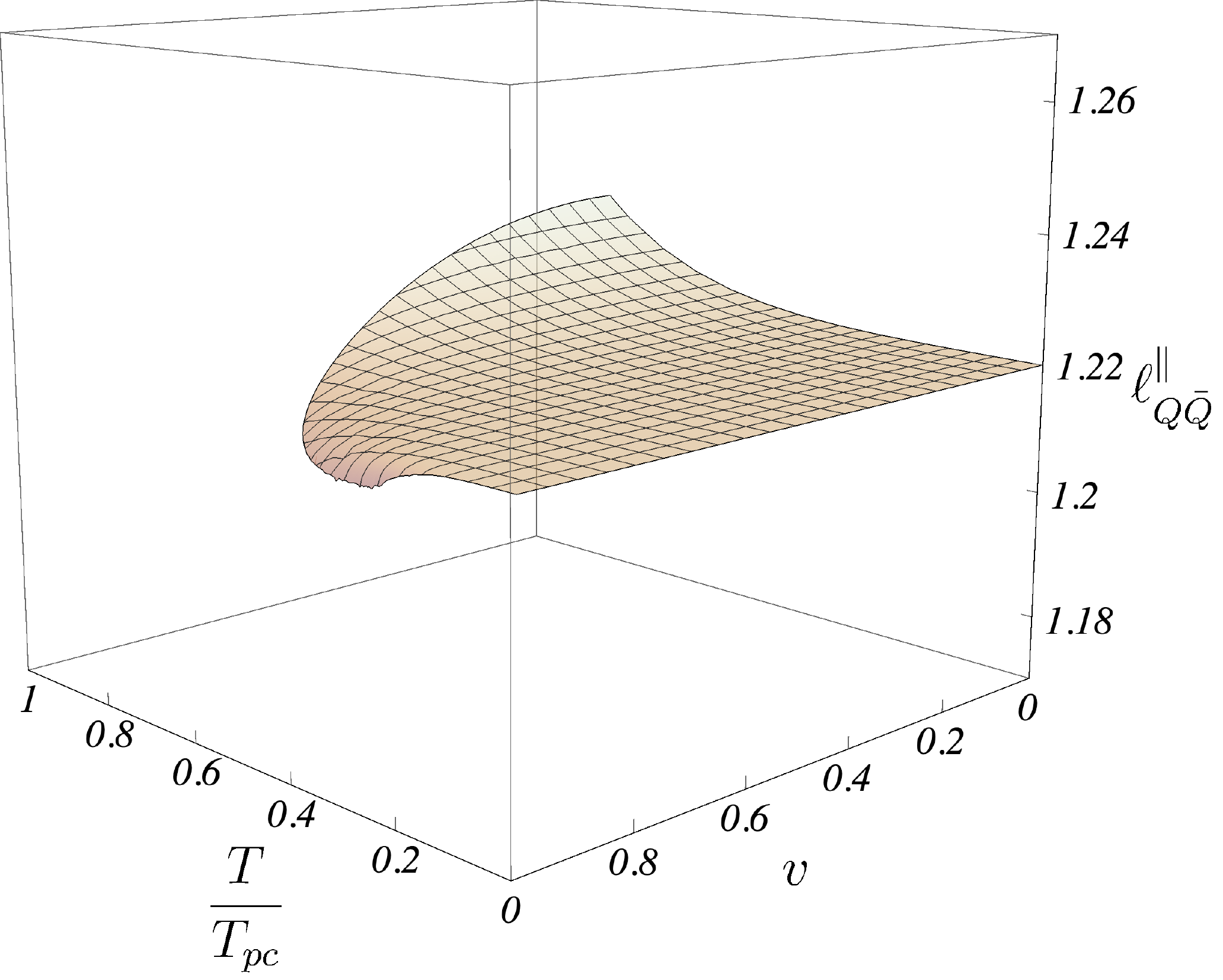}}}$
\caption{{\small $\boldsymbol{\ell}_{\QQb}(\text{fm})$ as a function of $v$ and $T$. The left and right panels refer respectively to the transverse and longitudinal cases.}}
\label{Lcs}
\end{figure}

In section 2 the formulas for the string breaking distance were derived using the linear approximation for $E_{\QQb}$ at large quark separation. To check if this approximation works well, we plot the energies in figure \ref{Elarge}, and we see that it does work.
One important observation one can make from the plots is that the string breaking distance slowly varies with temperature and velocity. Using the explicit expressions \eqref{lc-per}-\eqref{lc-par} together with \eqref{supp}, we can analyze this in more detail. In figure \ref{Lcs} we plot the string breaking distance vs temperature 
and velocity for both cases. In the transverse case, it is a slowly increasing function which remains approximately constant near the origin. Its deviation from the constant value ($\boldsymbol{\ell}_{\QQb}^{(0)}=1.22\,\text{fm}$) becomes visible for larger values of $T$ and $v$, and reaches a maximal value on the domain boundary, where it is of order $2\%$. At $T = 0$ the string breaking distance is independent of $v$, as expected due to Lorentz invariance. In the longitudinal case, it behaves in a more complicated way. For low velocities, the behavior is similar to that in the transverse case. Then at $v\approx 0.5$, the function begins to transform from increasing to decreasing and finally becomes decreasing for high velocities (except for the $v$-axis). The reason for such a different behavior is that the string tension $\estv$ is independent of velocity. As before, the deviation from the constant value doesn't exceed $2\%$.

\begin{figure}[htbp]
\centering
$\vcenter{\hbox{\includegraphics[width=6.9cm]{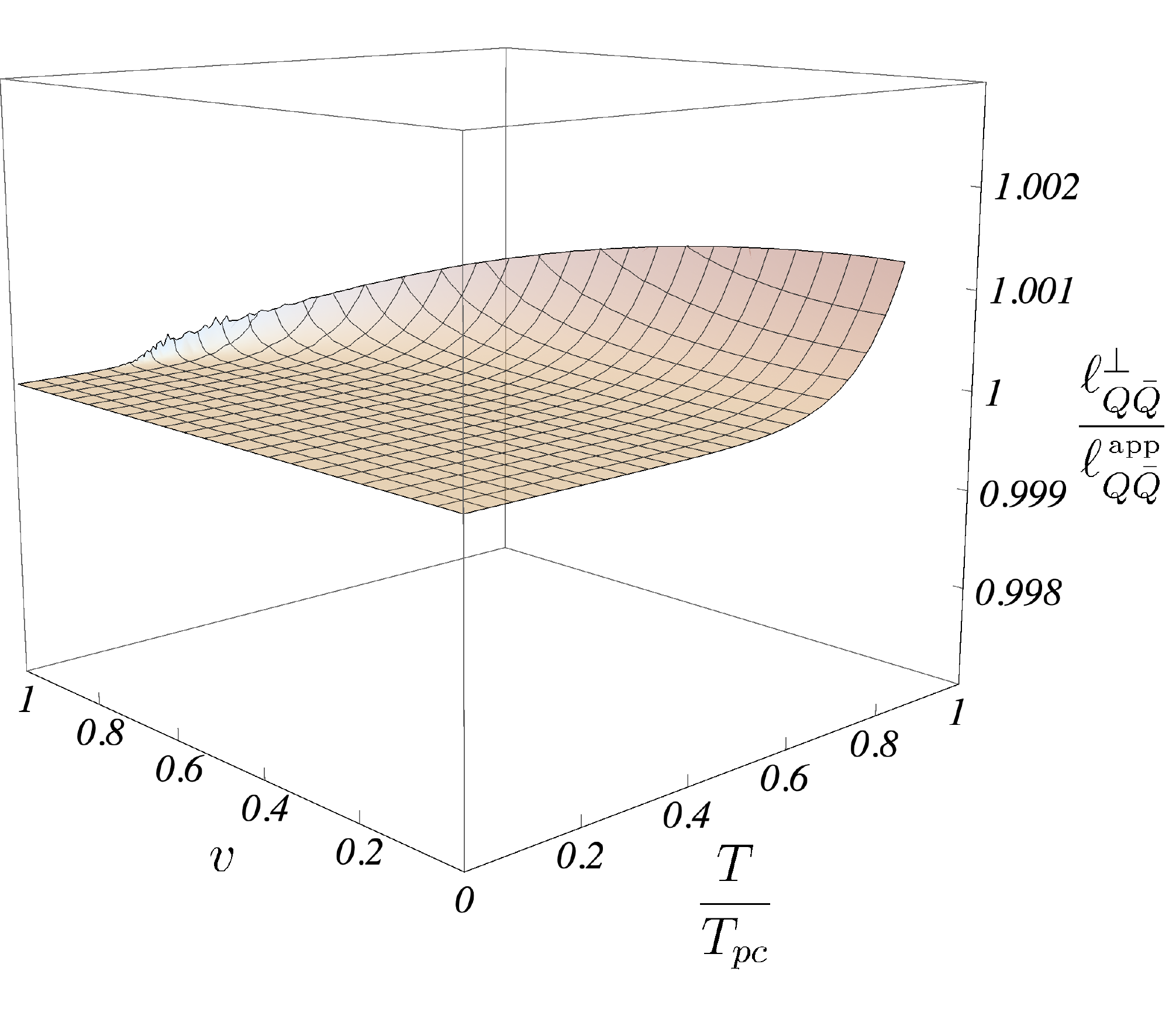}}}$
\hspace{0.5cm}
$\vcenter{\hbox{\includegraphics[width=7.6cm]{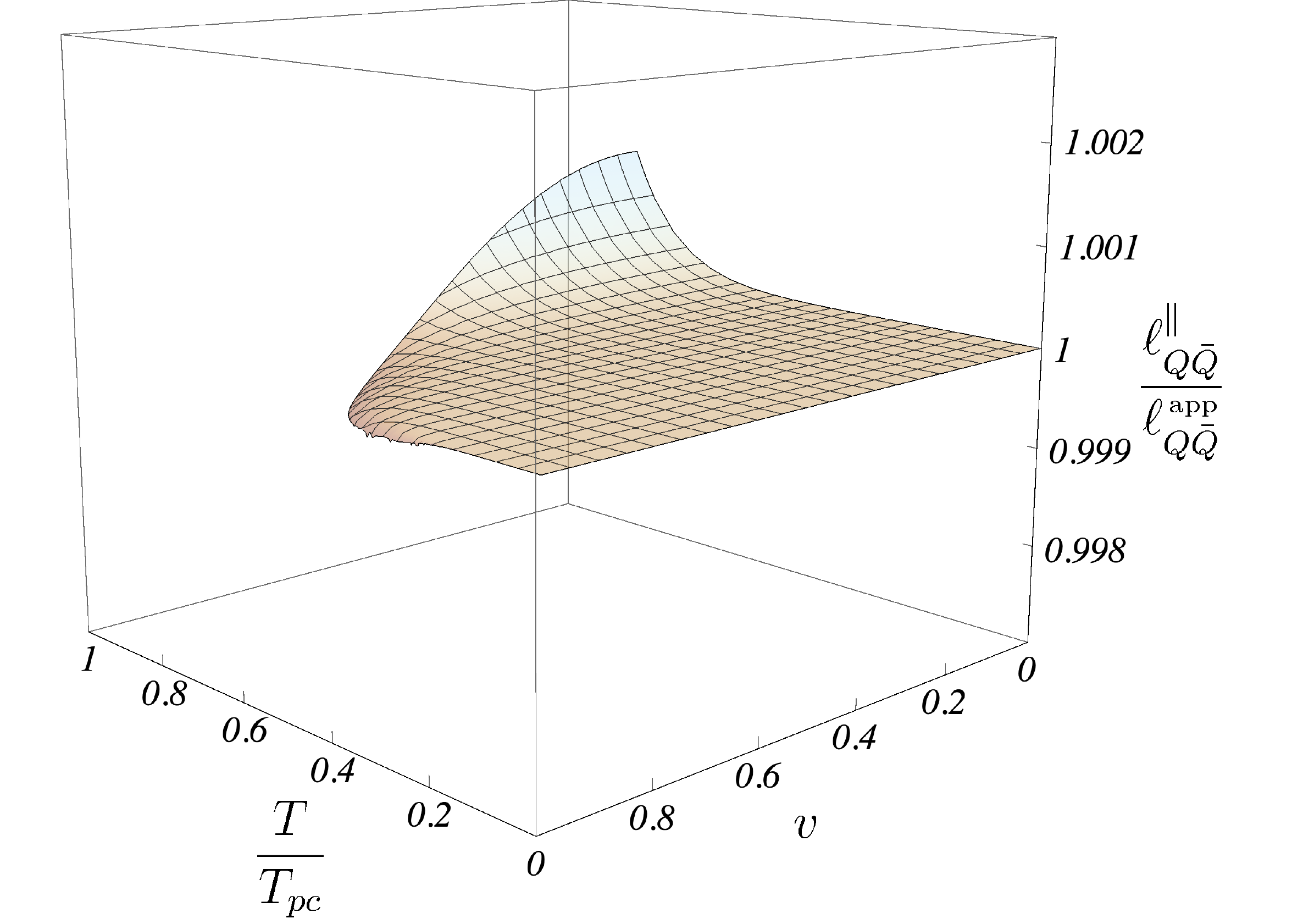}}}$
\caption{{\small $\frac{\boldsymbol{\ell}_{\QQb}}{\boldsymbol{\ell}_{\QQb}^{\text{\tiny app}}}$ ratio as a function of $T$ and $v$.}}
\label{approx-f}
\end{figure}

Both expressions for the string breaking distance look awkward for practical use. A reasonable approximation can be obtained by studying the low-temperature asymptotic expansion. The leading temperature correction to the string breaking distance $\boldsymbol{\ell}_{\QQb}^{(0)}$ is a sum of two terms such as $\gamma^2 T^4$ and $T^4$. So we have 

\begin{equation}\label{approx}
\boldsymbol{\ell}_{\QQb}^{\text{\tiny app}}
=
\boldsymbol{\ell}_{\QQb}^{(0)}	
\biggl(
1+a \gamma^2 \frac{T^4}{T_{pc}^4}+b \frac{T^4}{T^4_{pc}}
\,\biggr)
\,.
\end{equation}
A numerical calculation gives $a^{\bot}=0.026$ and $b^{\bot}=0$ for $\boldsymbol{\ell}_{\QQb}^{\bot}$, and $a^{\parallel}=-0.022$ and $b^{\parallel}=0.048$ for $\boldsymbol{\ell}_{\QQb}^{\parallel}$.\footnote{Note that $b^{\bot}=-0.0004$, with four digits after the decimal point.} To see how good this approximation is, we plot the ratio between $\boldsymbol{\ell}_{\QQb}$ and $\boldsymbol{\ell}_{\QQb}^{\text{\tiny app}}$ in figure \ref{approx-f}.
It is seen that it is indeed quite good. The maximal discrepancy between the two is less than $0.2\%$.

\section{Concluding Comments}
\renewcommand{\theequation}{4.\arabic{equation}}
\setcounter{equation}{0}

In our calculations we have assumed that the mass of $Q$ is much larger than that of $q$. Certainly, this is the case in the static limit where the $Q$'s are infinitely heavy. However in practice one is interested in the $c$ and $b$ quarks whose masses are large but still finite. It is interesting to see if the $u$ and $d$ quarks can be thought as light for the parameter values we are using. The estimate made in \cite{a-3Qb} gives $m_{u/d}=46.6\,\text{MeV}$. Thus, the $q$'s have the larger mass than the physical masses of $u$ and $d$. The reason for this is our fit to the lattice data of \cite{bulava} where the pion is twice heavier than in the real world. Nevertheless this estimate allows us to treat the $q$'s as light with respect to $c$ and $b$. It is worth mentioning that for this value of the pion mass lattice studies point towards a confinement-deconfinement phase transition \cite{bor}, as we also saw in section 3.  

The analysis that we made is valid deeply inside the hadronic phase. But when we consider the behavior near the critical line, namely in the region between the solid and dashed curves as shown in figure \ref{q}, the disconnected diagram with light quarks is not energetically favorable anymore and we should regard the disconnected configuration without light quarks as favorable. The physical meaning of this is that near the critical line the hadronic phase contains not only color-singlets, but besides that, some amount of color objects. In this case one can define the string breaking distance $\ell_{\Q}$ by equating $E_{\QQb}$ with $2E_{\Q}$ \cite{a-strb1}. It is a characteristic scale which corresponds to string breaking by thermal fluctuations. The corresponding analysis for the transverse and longitudinal cases proceeds in an obvious way. We will not make it here, instead we will restrict ourselves to the limiting case of zero velocity which enables us to quickly gain some insight into the problem.  

\begin{figure}[htbp]
\centering
\includegraphics[width=6cm]{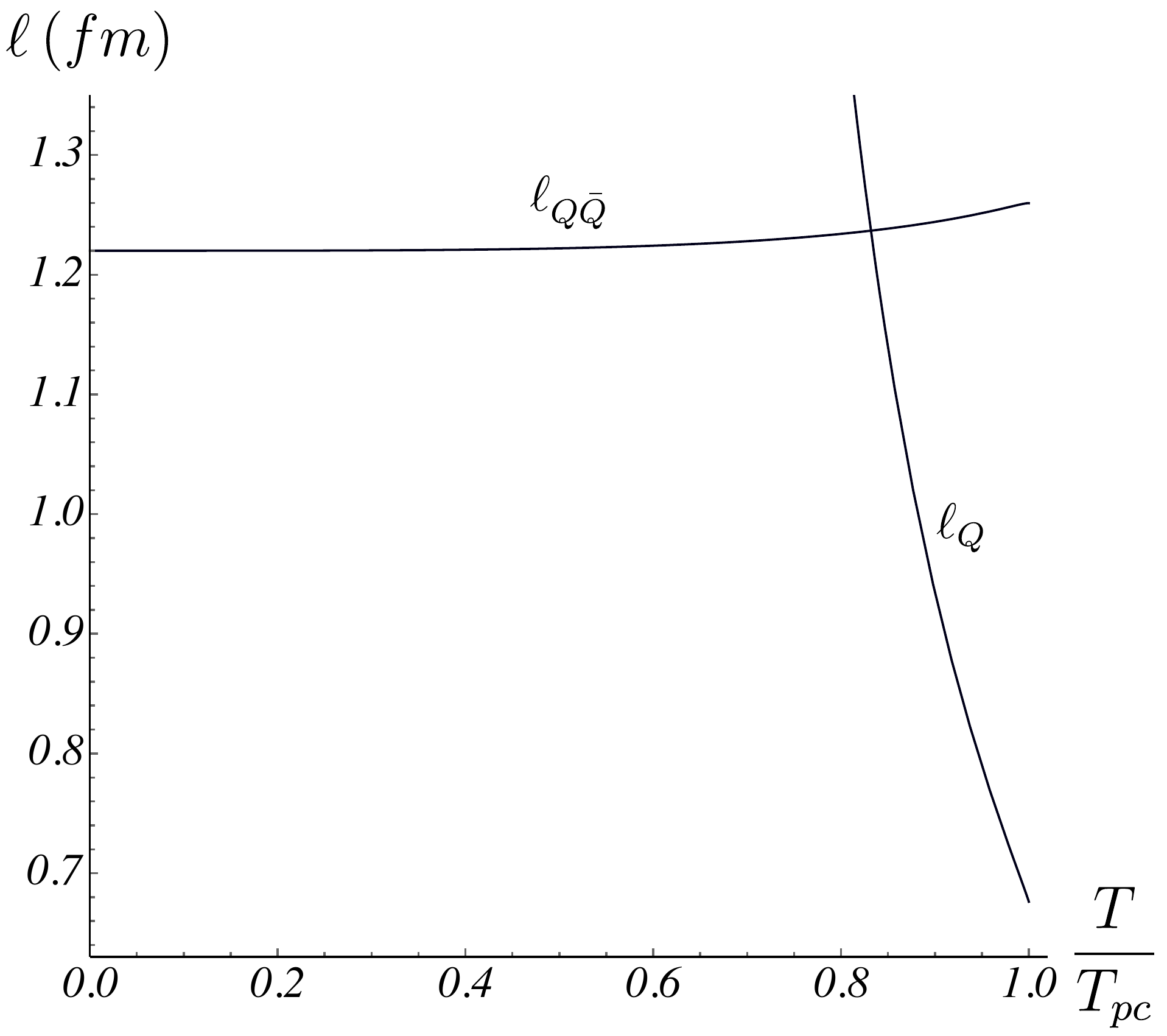}
\hspace{2.2cm}
\includegraphics[width=5.75cm]{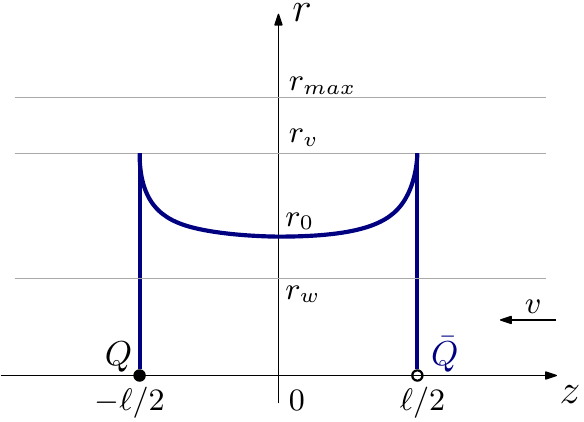}
\caption{{\small Left: String breaking distances vs temperature. Right: A spiky string configuration for $\rv<\rmax$.}}
\label{Lc-Q}
\end{figure}

From \eqref{lc-per}, the string breaking distance characterizing the decay mode $Q\bar Q\rightarrow Q\bar q+q\bar Q$ is 

\begin{equation}\label{lcT0}
\boldsymbol{\ell}_{\QQb}=
\frac{2}{\estv(\rmin)}
\biggl(\n\sqrt{f}\,\frac{\ep^{\frac{\s}{2}\rq^2}}{\rq}
+
\sqrt{\s}\Bigl(Q(q)-Q(\s\rmin^2)\Bigr) 
+
\int_0^{\rmin}\frac{dr}{r^2}\,\ep^{\s r^2}
\Bigl(1-\Bigl[1-\frac{\estv^2(\rmin)}{\estv^2}\Bigr]^{\oh}\Bigl)
\biggr)
	\,.
\end{equation}
On the other hand, the string breaking distance for the decay mode $Q\bar Q \rightarrow Q+\bar Q$ can be written as 

\begin{equation}\label{lcQ}
\boldsymbol{\ell}_{\Q}=
\frac{2}{\estv(\rmin)}
\biggl(
\sqrt{\s}\Bigl(Q(h)-Q(\s\rmin^2)\Bigr) 
+
\int_0^{\rmin}\frac{dr}{r^2}\,\ep^{\s r^2}
\Bigl(1-\Bigl[1-\frac{\estv^2(\rmin)}{\estv^2}\Bigr]^{\oh}\Bigl)
\biggr)
	\,.
\end{equation}
Now, specializing to the blackening factor \eqref{f-az}, we plot these as a function of temperature using the same parameter values as before. The result is presented in the left panel of figure \ref{Lc-Q}.
As we saw in section 3, at low temperatures the string breaking distance remains approximately constant and then, as temperature increases, it becomes weakly temperature dependent. The picture changes drastically when temperature passes a certain value $T_{\ast}$, which is determined by equation $E_{\Qqb}=E_{\Q}$.\footnote{In the model we are considering, $T_\ast\approx 0.830\,T_{pc}\approx110\,\text{MeV}$ at $\s=0.450\,\text{GeV}^2$.} In the interval $T_\ast\leq T\leq T_{pc}$ the string breaking distance steeply decreases with increasing temperature. Nevertheless, the linear approximation for $E_{\QQb}$ still holds as the smallest value of $\ell_{\Q}$ is only about $0.66\,\text{fm}$ (see figure \ref{Elarge}). The question arises of whether a similar approach works for higher temperatures and if so, how the so defined scale is related to the Debye screening length. We do not know the answer to that question. One trouble is that in the string models the connected configuration at high temperature exists only for small values of $\ell$ so that quark separation does not exceed a certain critical value \cite{djg}.

We conclude by discussing subdominant string configurations. These were discussed in \cite{a-strb1} at zero wind velocity. The conclusion is that they are important for understanding excited states, but have negligible effect on the ground state. One may wonder, what happens if the wind blows? A partial answer is that this conclusion is unaffected by the wind, at least for the configurations of \cite{a-strb1}. A novel string configuration which could change it is sketched in the right panel of figure \ref{Lc-Q}. An important point is that such a spiky configuration exists only in the presence of the wind. A convenient way of thinking about it is as follows. Take a disconnected string configuration as that shown in the right panel of figure \ref{disc}. Because the string tension at the induced horizon is zero, the strings are stable. Then attach another string to those by connecting the endpoints at $r=\rv$. If $\rv<\rmax$, then the attached string hangs down towards the boundary. Like the strings of figure \ref{conconfs}, it approaches the soft wall as quark separation increases. This implies that the energies of the configurations are equal at leading order in $\ell$. What about the next order? To answer this question, we need to analyze the spiky configuration in more detail.

As an example, let us briefly consider the longitudinal case shown in Figure \ref{Lc-Q}. The side strings have already been discussed in subsection 2.3. The string attached to them may be analyzed similarly as described in subsection 2.2. So, we take the static gauge $\xi_1=t$ and $\xi_2=z$, and consider $r$ as a function of $z$. The boundary conditions at the string endpoints are now $r(\pm\ell/2)=\rv$. The Nambu-Goto action has the same form as that in \eqref{NGp}. Hence a first integral of the equation of motion is given by \eqref{I-par}. As usual, we set $I=\estv(\r0)$, where $\r0$ is a turning point which now corresponds to a local minimum (see the figure above). The string length along the $z$-axis is then obtained by integrating the first integral. We have

\begin{equation}\label{l-spike}
\ell=2\int_{\r0}^{\rv} \frac{dr}{f}\,\sqrt{\fv} 
\biggl[\,
\frac{\estv^2}{I^2}-1\,
\biggr]^{-\oh}
\,,
\end{equation}
with a symmetry factor of $2$.

Since the string is static, its energy is computed directly from the action \eqref{NGp}. The key simplification is that regularization is not needed. The reason for this is that the string does not approach the boundary, and as a result, the effective string tension $\estv$ remains finite. After adding the obtained result to those for the side strings, we can write the total energy as

\begin{equation}\label{E-spike}
E_{\QQb}=
2E_{\Q}
+
2\g\int_{\r0}^{\rv} dr \frac{\est}{\sqrt{f}}
\biggl[\,1-\frac{I^2}{\estv^2}\,\biggr]^{-\oh}
\,,
\end{equation}
where $E_{\Q}$ is given by \eqref{EQ}. Thus, the energy of the configuration is written in parametric form as $E_{\QQb}=E_{\QQb}(\r0)$ and $\ell=\ell(\r0)$. The parameter takes values on the interval $[\rmin ,\rv]$. A simple analysis shows that large $\ell$'s correspond to the region near the lower endpoint $\r0=\rmin$ and the long distance behavior of $E_{\QQb}$ is 

\begin{equation}\label{Elarge-spike}
	E_{\QQb}=\sigma\ell
	+
	2E_{\Q}
	+
	2\g\int_{\rmin}^{\rv} dr \frac{\est}{\sqrt{f}}
	\biggl[\,1-\frac{\estv^2(\rmin)}{\estv^2}\,\biggr]^{\oh}
+o(1)
	\,,
\end{equation}
with the same $\sigma$ as in \eqref{Elarge-par}. 

Having found the asymptotic expansion, we can now compare it with that of section 3. The linear terms are the same, as we discussed above, but the constant terms are different.  The difference between those is\footnote{We subtract the constant term in the expansion \eqref{Elarge-par} from that in \eqref{Elarge-spike}.}

\begin{equation}\label{gap}
\frac{\Delta}{2\g}=
\int_{\rmin}^{\rv} dr \frac{\est}{\sqrt{f}}
	\biggl(
	1+\biggl[\,1-\frac{\estv^2(\rmin)}{\estv^2}\,\biggr]^{\oh}
	\biggr)
	+
	\int_{0}^{\rmin} dr \frac{\est}{\sqrt{f}}
	\biggl(
	1
	-
	\biggl[\,1-\frac{\estv^2(\rmin)}{\estv^2}\,\biggr]^{\oh}
	\biggr)
\,.
\end{equation}
The integrals are well-defined and positive. The latter follows from the fact that both integrands are non-negative. Thus, $\Delta>0$ and therefore the spiky configuration has no effect on string breaking related to the ground state. In a similar way, one can study the spiky string configuration in the transverse case and its relevance for string breaking. 

\vspace{.1cm}
{\bf Acknowledgments}

\vspace{.15cm}
\noindent This research is supported in part by RFBR Grant 18-02-40069. 


\end{document}